\documentclass{PoS}

\usepackage{lipsum}
\usepackage{amsmath}
\usepackage{microtype}

\title{EPPS16 -- Bringing nuclear PDFs to the LHC era}

\ShortTitle{EPPS16 -- Bringing nuclear PDFs to the LHC era}

\author{Kari J. Eskola\\
       University of Jyvaskyla, Department of Physics, P.O. Box 35, FI-40014 University of Jyvaskyla, Finland \\
       Helsinki Institute of Physics, P.O. Box 64, FI-00014 University of Helsinki, Finland \\
       E-mail: \email{kari.eskola@jyu.fi}}

\author{\speaker{Petja Paakkinen}
        \\ University of Jyvaskyla, Department of Physics, P.O. Box 35, FI-40014 University of Jyvaskyla, Finland \\
        E-mail: \email{petja.paakkinen@jyu.fi}}

\author{Hannu Paukkunen\\
       University of Jyvaskyla, Department of Physics, P.O. Box 35, FI-40014 University of Jyvaskyla, Finland \\
       Helsinki Institute of Physics, P.O. Box 64, FI-00014 University of Helsinki, Finland \\
       E-mail: \email{hannu.paukkunen@jyu.fi}}

\author{Carlos A. Salgado\\
       Instituto Galego de F\'{\i}sica de Altas Enerx\'{\i}as (IGFAE), Universidade de Santiago de Compostela, E-15782 Galicia, Spain \\
       E-mail: \email{carlos.salgado@usc.es}}

\abstract{We report on EPPS16, the first global analysis of nuclear parton distribution functions (nPDFs) to include LHC data. Also for the first time, a full flavour dependence of nPDFs is allowed. While the included Z and W data are found to have insufficient statistics to yield stringent constraints, the CMS 5.02 TeV proton--lead dijet data prove crucial in setting the shape of nuclear gluon modifications. With these and other observables being measured in proton--lead runs, we are experiencing a shift of nPDFs to the LHC precision era.}

\FullConference{
12th International Workshop on High-pT Physics in the RHIC/LHC Era\\
		2-5 October, 2017\\
		University of Bergen, Bergen, Norway}

\begin{document}

\section{Introduction}

With the data coming available from the LHC proton--lead runs, nuclear parton distribution functions (nPDFs) need to adjust to this LHC precision era. We report here on EPPS16~\cite{Eskola:2016oht}, the first nPDF global analysis to include LHC data, discussing the impact of these data on the nPDFs and their uncertainties. A major improvement is also a fully flavour-dependent parametrization of the light-parton nuclear modifications, yielding less biased uncertainty estimates. Here, the neutrino--nucleus deep inelastic scattering data is important in constraining the modifications of individual quark flavours.

\section{Parametrization}

In EPPS16, the PDFs of a proton inside a nucleus $A$, $f_i^{{\rm p}/A}(x,Q^2)$, with the index $i$ labeling the parton species, are obtained by multiplying the free proton PDFs $f_i^{{\rm p}}(x,Q^2)$ with nuclear modification functions $R_i^A(x,Q^2)$,
\begin{equation}
  f_i^{{\rm p}/A}(x,Q^2) = R_i^A(x,Q^2) f_i^{{\rm p}}(x,Q^2),
\end{equation}
and the neutron PDFs $f_i^{{\rm n}/A}(x,Q^2)$ are obtained from these using isospin symmetry. The free proton PDFs used in this analysis are those of CT14~\cite{Dulat:2015mca}. The PDFs of a full nucleus are then constructed with
\begin{equation}
  f_i^A(x,Q^2) = \frac{Z}{A} f_i^{{\rm p}/A}(x,Q^2) + \frac{N}{A} f_i^{{\rm n}/A}(x,Q^2).
\label{eq:fullnuc}
\end{equation}

\begin{floatingfigure}[l]
  \begin{minipage}{0.48\textwidth}
    \includegraphics[width=\textwidth]{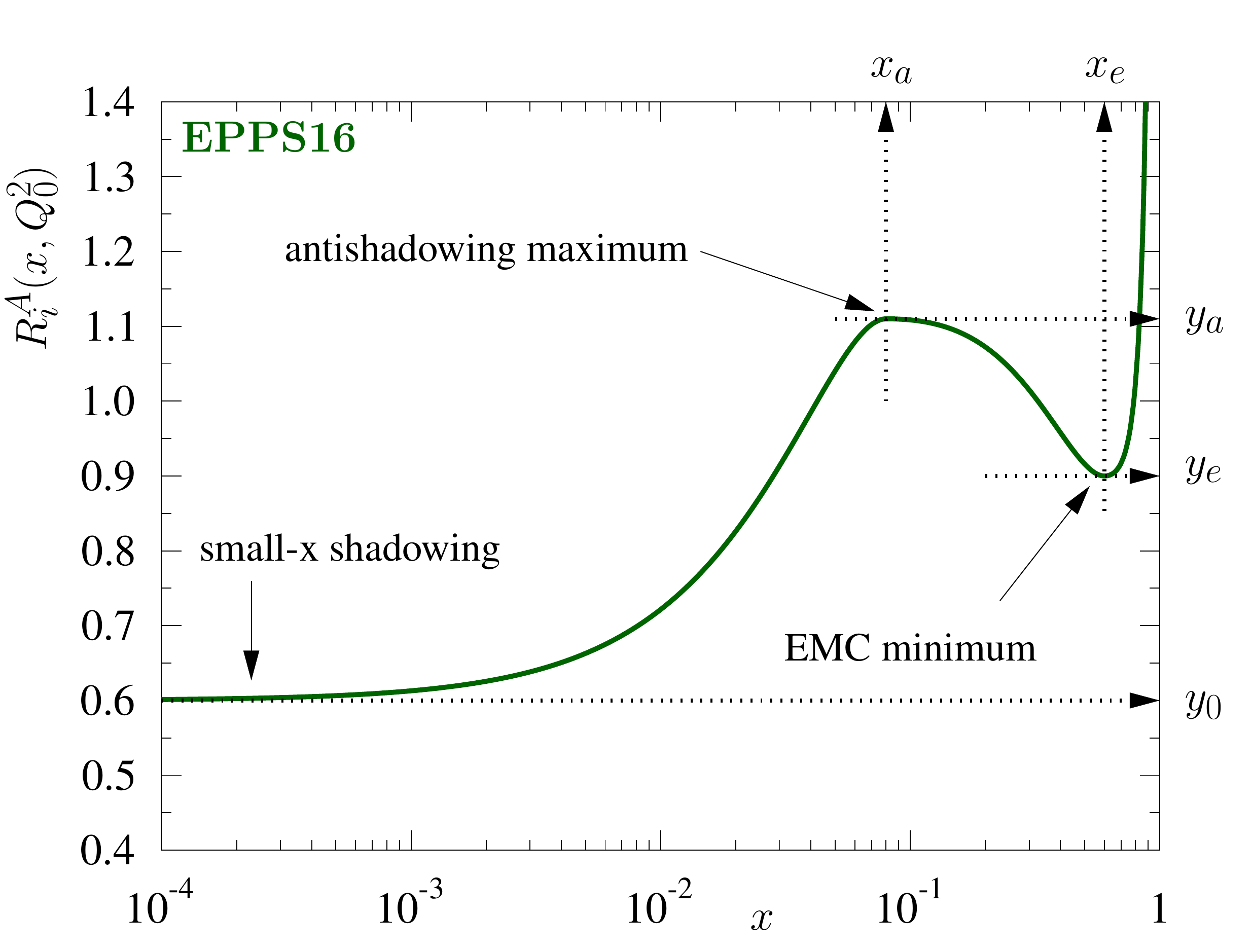}
  \end{minipage}
  \caption{Illustration of the EPPS16 parametrization of the nuclear modification functions. Figure from Ref.~\cite{Eskola:2016oht}.}
  \label{fig:param}
\end{floatingfigure}

The modification functions are para\-met\-ri\-zed at an initial scale $Q^2_0 = m^2_{\rm charm}$ with a piecewise function illustrated in Figure~\ref{fig:param}. Parameters $y_0$, $y_a$ and $y_e$ govern the amount of small-$x$ shadowing, antishadowing maximum and EMC minimum, respectively, while $x_a$ and $x_e$ control the $x$ locations of the latter two. The mass-number dependence of $y_i$ are para\-met\-ri\-zed with
\begin{equation}
  y_i(A) = y_i(A_{\rm ref}) \left(\frac{A}{A_{\rm ref}} \right)^{\gamma_i \left[y_i(A_{\rm ref}) - 1\right]}, \quad \gamma_i \geq 0,
\end{equation}
such that larger nuclei exhibit larger modifications, using $A_{\rm ref} = 12$. The deuteron ($A=2$) is considered to be free from nuclear effects.

In EPPS16 we were able for the first time to parametrize the full flavour dependence of the light-parton nuclear modifications,
\begin{equation}
  R_{\textcolor{black}{u_{\rm V}}}^A(x,Q^2_0) \neq R_{\textcolor{black}{d_{\rm V}}}^A(x,Q^2_0),
  \qquad
  R_{\textcolor{black}{\bar{u}}}^A(x,Q^2_0) \neq R_{\textcolor{black}{\bar{d}}}^A(x,Q^2_0) \neq R_{\textcolor{black}{\bar{s}}}^A(x,Q^2_0).
  \label{eq:flavoursym}
\end{equation}
This is an important improvement compared to earlier analyses, which have relied on simplifying assumptions and thus are more susceptible to a bias. Independent valence parametrizations were used also in the nCTEQ15 analysis~\cite{Kovarik:2015cma}; we will present a comparison in Section~\ref{sec:comp}.

\section{Data treatment}

\begin{floatingfigure}[r]
  \begin{minipage}{0.48\textwidth}
    \includegraphics[width=\textwidth]{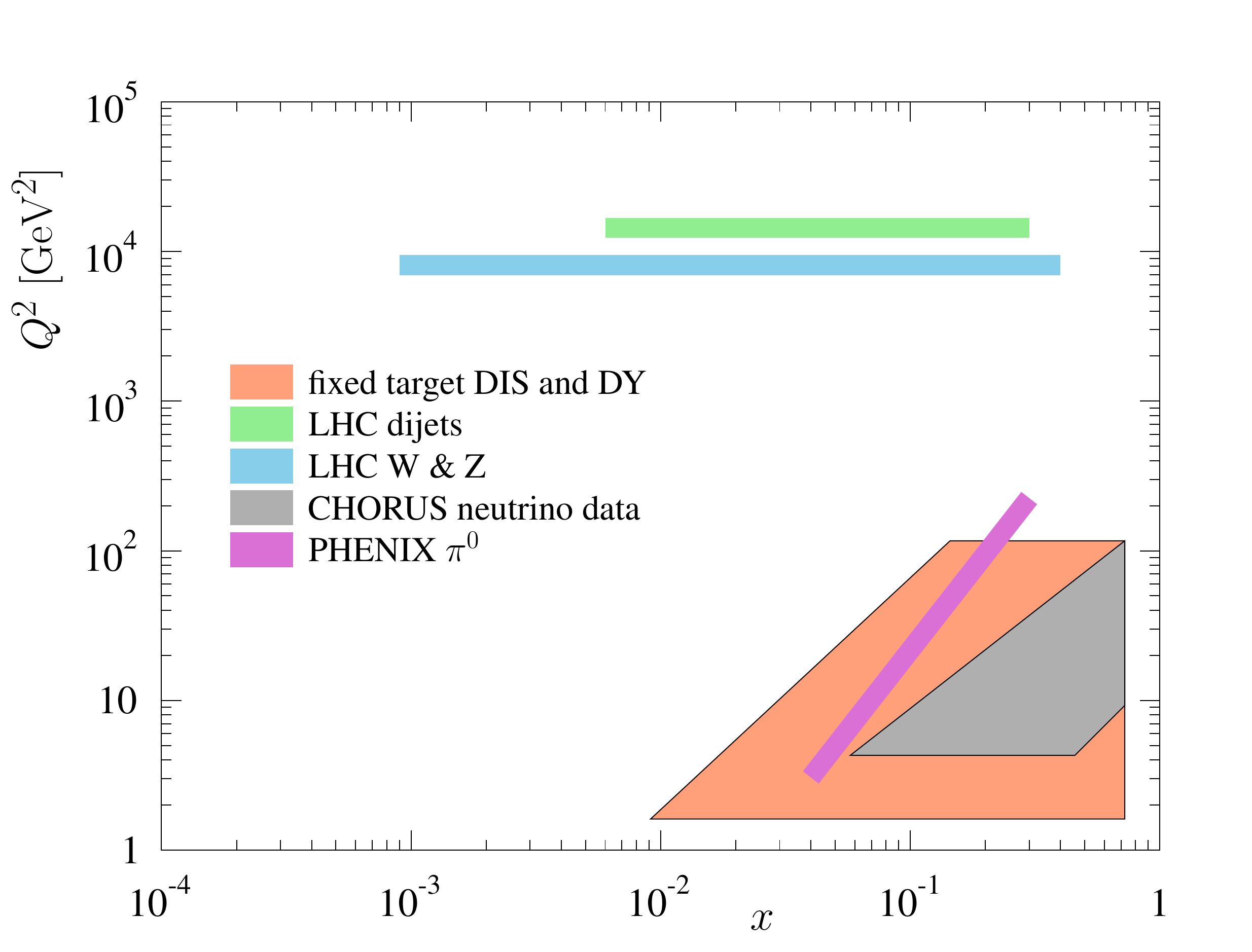}
  \end{minipage}
  \caption{Kinematic reach of the data used in the EPPS16 analysis. Figure from Ref.~\cite{Eskola:2016oht}.}
  \label{fig:xQ2}
\end{floatingfigure}

In Figure~\ref{fig:xQ2} we show the kinematic reach of the data used in the EPPS16 analysis. As in the earlier EPS09 analysis~\cite{Eskola:2009uj}, we use deep inelastic scattering~(DIS) and proton--nucleus Drell--Yan~(DY) dilepton-production data from older fixed-target experiments, as well as pion-production data from RHIC. These are included in the form of nuclear ratios, reducing experimental systematic uncertainties and dependence on choice of the baseline free proton PDFs.
As data in this form are not available for the LHC observables and neutrino--nucleus DIS used in EPPS16, other kind of measures need to be taken.

We include the LHC data as forward-to-backward ratios and the neutrino--nucleus DIS data as normalized cross sections,
propagating also the correlated systematic uncertainties to the normalized cross sections. For the pion--nucleus DY data, we use either cross-section ratios of different target nuclei or different projectile pion charges, both of these observables being insensitive to pion PDFs~\cite{Paakkinen:2016wxk}. Now that we allow flavour separation in nuclear modifications, it is also important to recover the true charged-lepton DIS structure functions from the published ``isoscalarized'' ones, see Ref.~\cite{Eskola:2016oht} for details.

\section{Impact of LHC data}

\begin{floatingfigure}[t]
  \hspace{-0.18cm}
  \begin{minipage}{0.333\textwidth}
    \includegraphics[width=\textwidth]{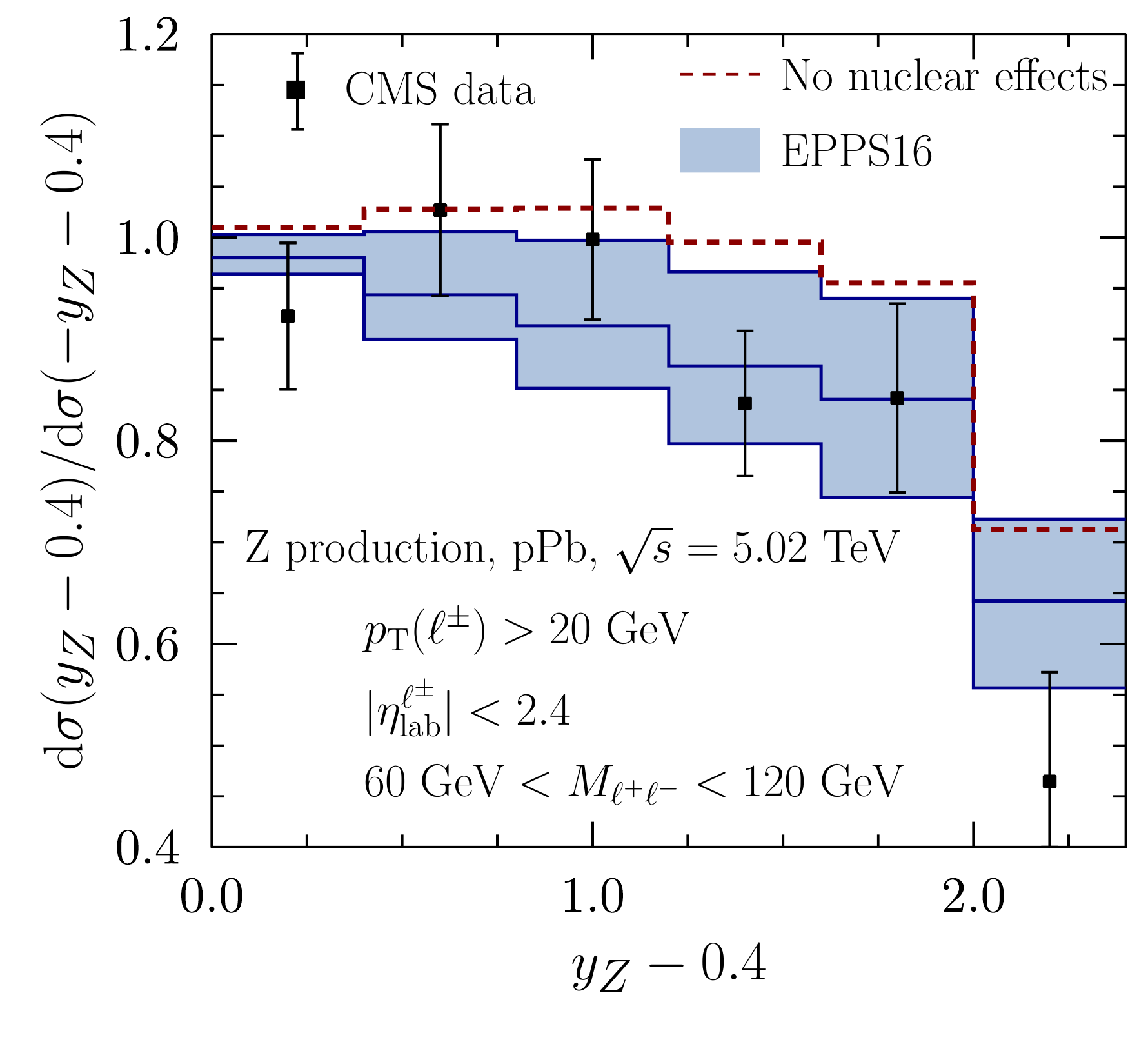}
    \includegraphics[width=\textwidth]{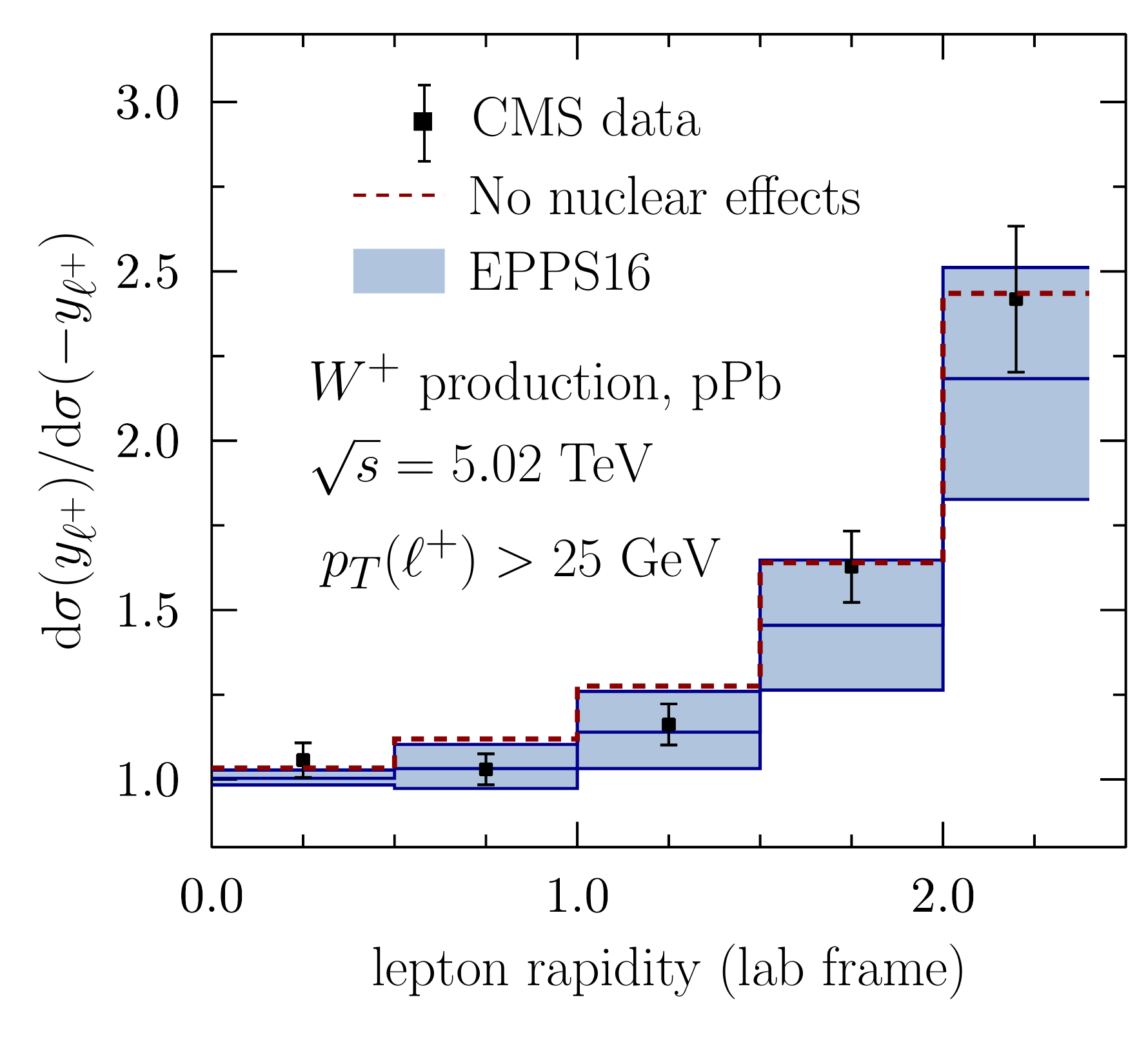}
  \end{minipage}
  \hspace{-0.18cm}
  \begin{minipage}{0.333\textwidth}
    \includegraphics[width=\textwidth]{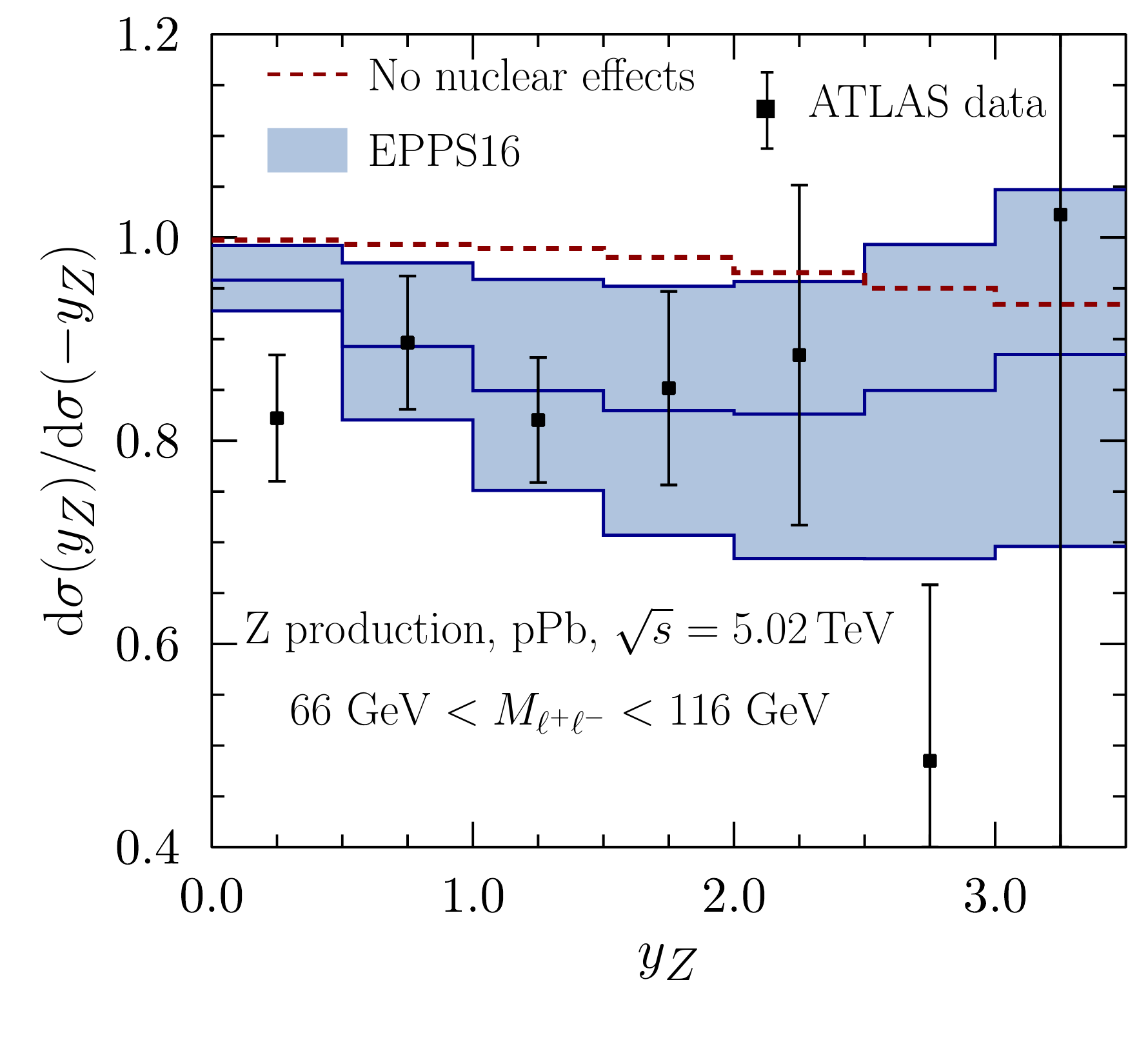}
    \includegraphics[width=\textwidth]{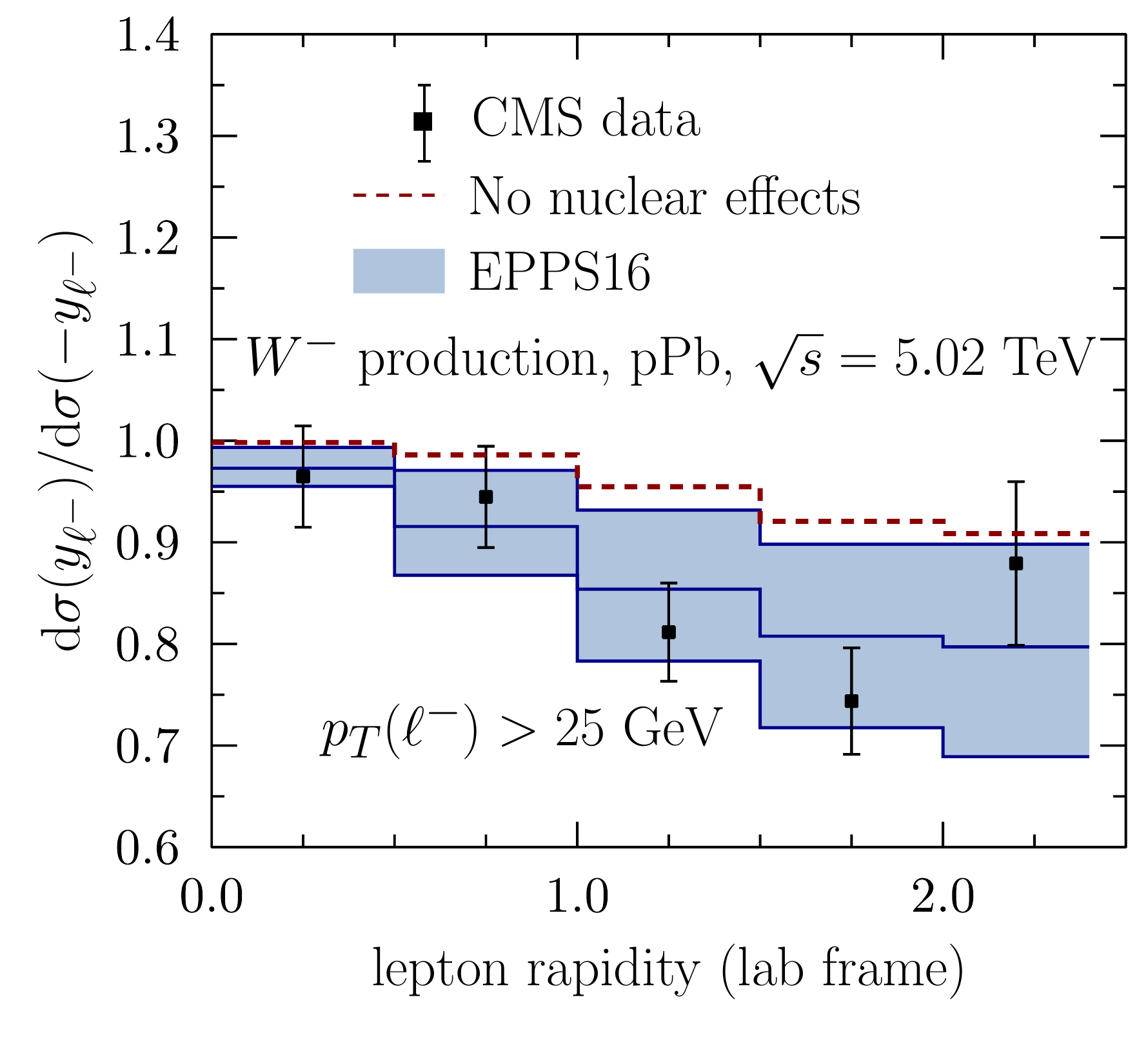}
  \end{minipage}
  \hspace{-0.18cm}
  \begin{minipage}{0.333\textwidth}
    \includegraphics[width=\textwidth]{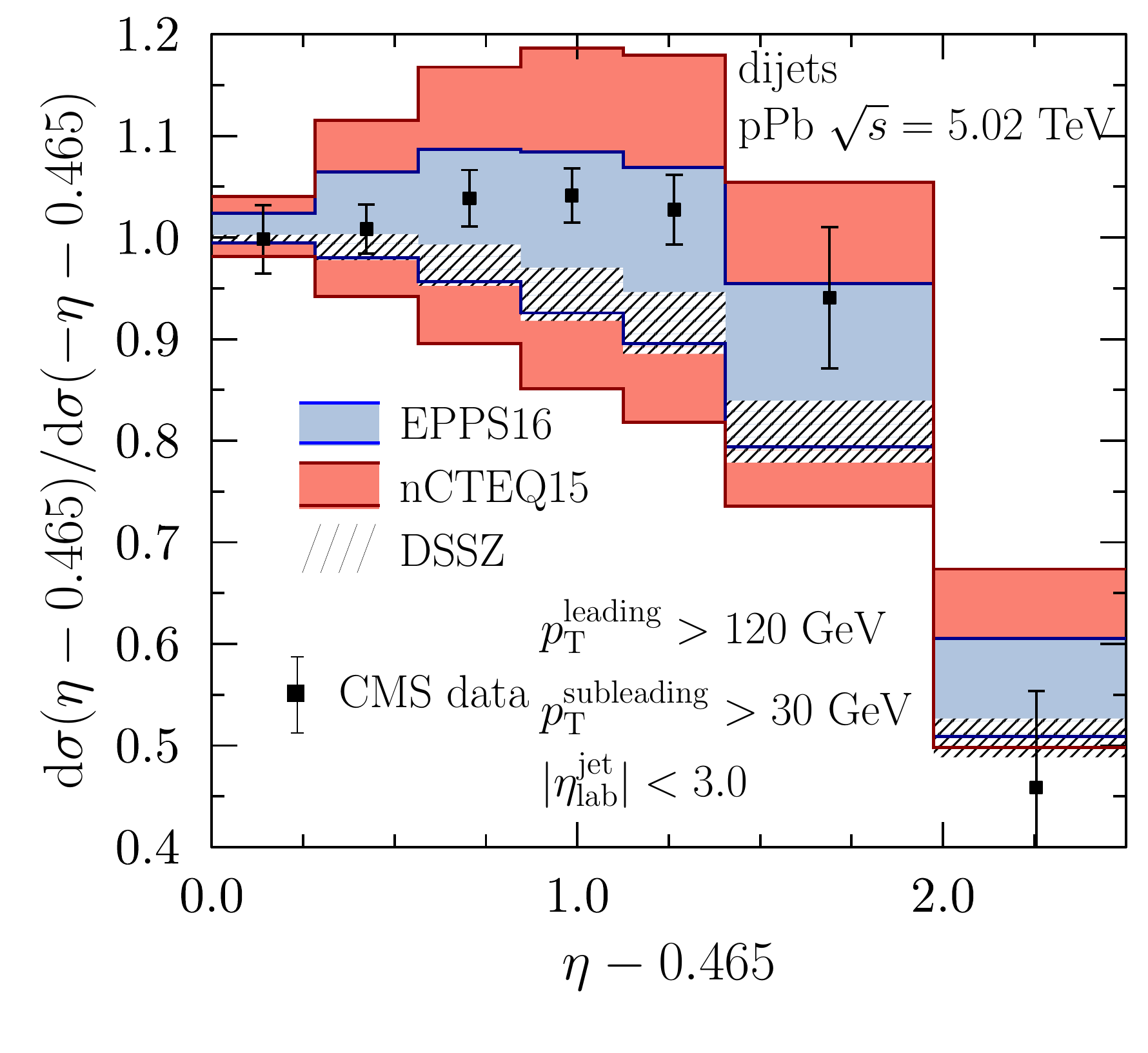}
  \end{minipage}
  \hspace{-0.18cm}
  \vspace{-1ex}
  \caption{The LHC 5.02 TeV proton--lead data on Z production~\cite{Khachatryan:2015pzs,Aad:2015gta} (upper left and middle panels), W production~\cite{Khachatryan:2015hha} (lower left and middle panels) and dijets~\cite{Chatrchyan:2014hqa} (rightmost panel) compared with nuclear PDFs. Figures from Ref.~\cite{Eskola:2016oht}.}
  \label{fig:LHC}
\end{floatingfigure}

The LHC data included in the EPPS16 analysis are presented in Figure~\ref{fig:LHC}. The Z production data from CMS~\cite{Khachatryan:2015pzs} and ATLAS~\cite{Aad:2015gta} clearly deviate from a prediction with no nuclear effects and are in concordance with the fit obtained in EPPS16. This supports net nuclear shadowing at small $x$, but the constraints from these data are limited due to low statistics. Also, since correlations on the systematic errors were not available, the experimental uncertainties were added in quadrature when forming the forward-to-backward ratios, undermining the constraining power of these data.
In the case of the CMS W production data~\cite{Khachatryan:2015hha}, forward-to-backward ratios were directly published, and accordingly the experimental uncertainties are smaller. Still, with only 10 data points, the relative weight of these data is rather small in the global fit, and more data are needed for better constraints. Again, a good fit is obtained in EPPS16.

\begin{floatingfigure}[t]
  \includegraphics[width=\textwidth]{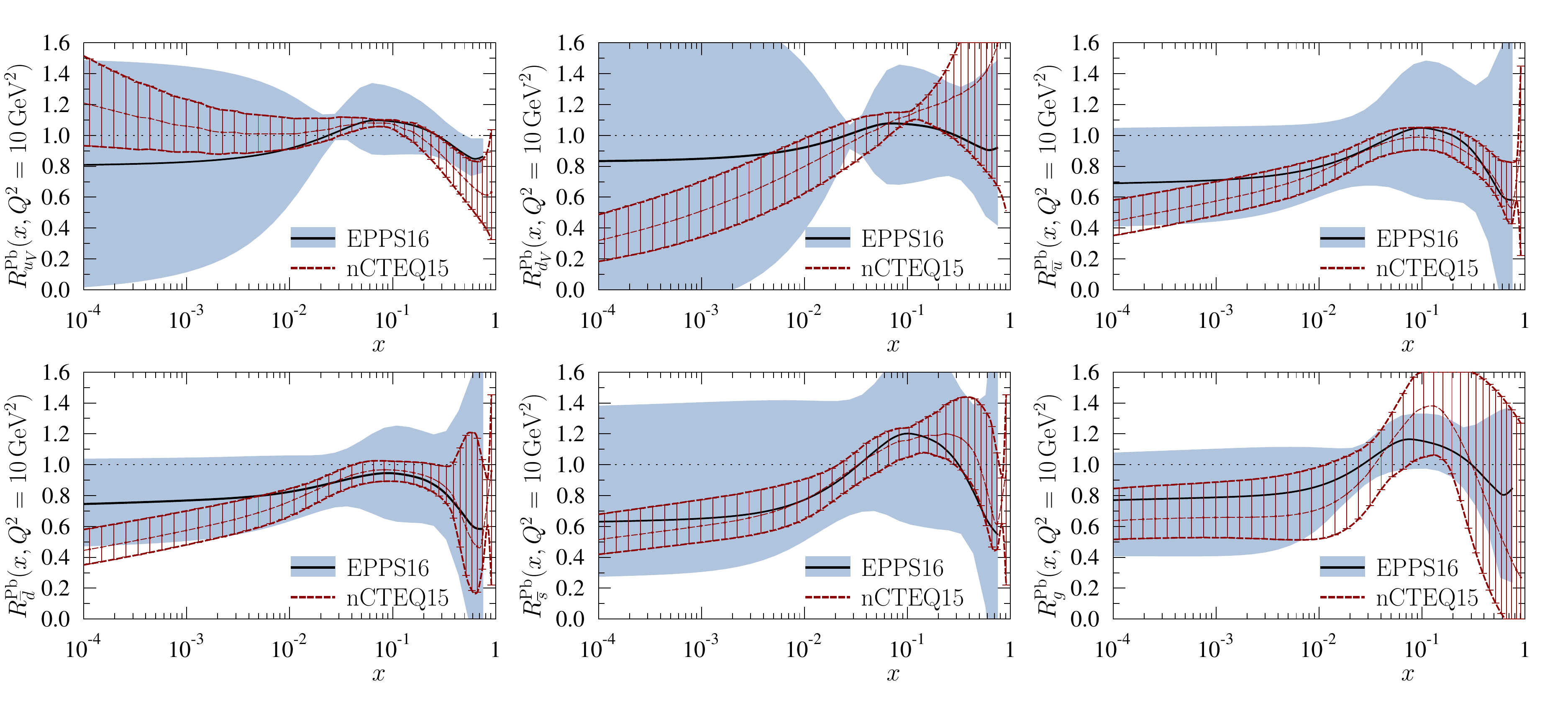}
  \vspace{-4ex}
  \caption{Nuclear modification functions for lead at $Q^2 = 10\ {\rm GeV}^2$ from the EPPS16 and nCTEQ15 analyses. Figure from Ref.~\cite{Eskola:2016oht}.}
  \label{fig:nPDFs}
\end{floatingfigure}

While the impact of the electroweak observables is small with the present statistics, important constraints for gluons are obtained from the CMS dijet data~\cite{Chatrchyan:2014hqa}. If these data were not included, the obtained central prediction for gluon nuclear modification in EPPS16 would exhibit no EMC suppression, but once these data are taken into account, a clear EMC slope is obtained (see Figure~\ref{fig:nPDFs}). Correspondingly, the gluon uncertainties at mid-to-high $x$ are significantly reduced when the dijet data are included.

\section{Results and comparison}
\label{sec:comp}

Results of the EPPS16 analysis for nuclear modifications in lead at $Q^2 = 10\ {\rm GeV}^2$ are shown in Figure~\ref{fig:nPDFs}. The uncertainties are obtained using the standard Hessian method with ``90\% confidence criterion'', see Ref.~\cite{Eskola:2016oht} for details. We note the following: Even though independently parametrized, the valence modifications appear very similar. This is primarily due to the inclusion of CHORUS neutrino--nucleus DIS data~\cite{Onengut:2005kv}. Without these data, the fit converges to a parameter region where $R_{{u_{\rm V}}}^A(x,Q^2_0)$ and $R_{{d_{\rm V}}}^A(x,Q^2_0)$ differ significantly from each other, but once these data are included, we find a preference for similar modifications. Comparing to results of the nCTEQ15 analysis, also shown in Figure~\ref{fig:nPDFs}, we find that they arrive with very different valence modifications, most likely since they do not include neutrino DIS and due to using isoscalarized charged-lepton DIS data.

\begin{floatingfigure}[r]
  \begin{minipage}{0.39\textwidth}
    \hspace{0.041\textwidth}\includegraphics[width=0.959\textwidth]{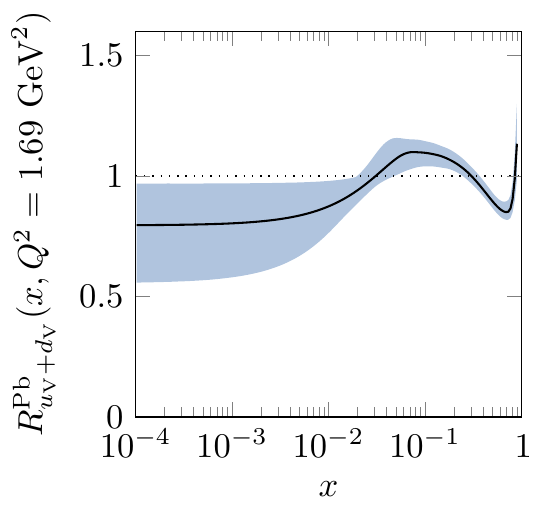} \\
    \hspace{0.0\textwidth}\includegraphics[width=\textwidth]{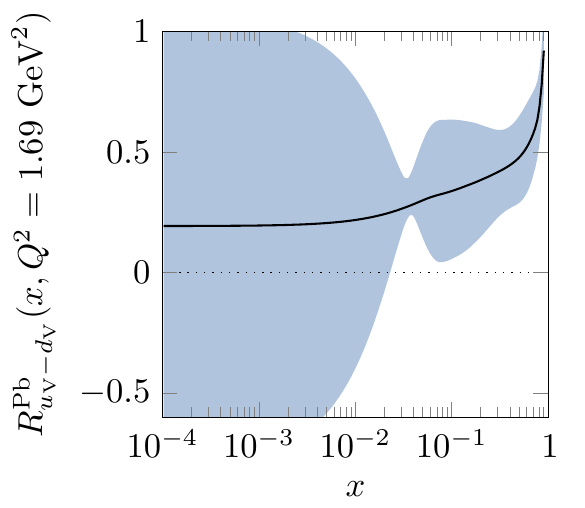}
  \end{minipage}
  \caption{Isospin decomposition of the EPPS16 valence quark PDFs in a lead nucleus.}
  \label{fig:avediff}
\end{floatingfigure}

We also find the EPPS16 valence uncertainties to be rather large, which might appear surprising as the included DIS data should constrain these well. This can be understood as follows: We can decompose the valence distributions of a full nucleus as
\begin{equation}
  \begin{split}
    f^A_{u_{\rm V}} = \left( R^A_{u_{\rm V} + d_{\rm V}} - \frac{A - 2Z}{A} R^A_{u_{\rm V} - d_{\rm V}} \right) \frac{f^p_{u_{\rm V}} + f^p_{d_{\rm V}}}{2}, \\
    f^A_{d_{\rm V}} = \left( R^A_{u_{\rm V} + d_{\rm V}} + \frac{A - 2Z}{A} R^A_{u_{\rm V} - d_{\rm V}} \right) \frac{f^p_{u_{\rm V}} + f^p_{d_{\rm V}}}{2},
  \end{split}
  \label{eq:decomp}
\end{equation}
where
\begin{equation}
  R^A_{u_{\rm V} + d_{\rm V}} = \frac{f^{p/A}_{u_{\rm V}} + f^{p/A}_{d_{\rm V}}}{f^p_{u_{\rm V}} + f^p_{d_{\rm V}}}, \quad R^A_{u_{\rm V} - d_{\rm V}} = \frac{f^{p/A}_{u_{\rm V}} - f^{p/A}_{d_{\rm V}}}{f^p_{u_{\rm V}} + f^p_{d_{\rm V}}}.
  \label{eq:isosp}
\end{equation}
Even for heavier nuclei like lead the amount of neutron excess is quite small, $\frac{A - 2Z}{A} \approx 0.2$. Hence when we construct any cross section out of the PDFs in Eq.~\eqref{eq:decomp}, they will be predominantly sensitive to the average modification $R^A_{u_{\rm V} + d_{\rm V}}$. Indeed, comparing the two isospin components of Eq.~\eqref{eq:isosp} in Figure~\ref{fig:avediff}, we find the EPPS16 uncertainties for $R^{\rm Pb}_{u_{\rm V} - d_{\rm V}}$ to be much larger than those in $R^{\rm Pb}_{u_{\rm V} + d_{\rm V}}$. To better constrain the difference, we would need more high-precision data on non-isoscalar nuclei.

Regarding the sea quarks, we find again all flavours to have rather similar modifications, this time also in accordance with nCTEQ15, where no flavour freedom was allowed. Also here the EPPS16 uncertainties are large due to allowing flavour freedom, with only the average being well constrained. Since nCTEQ15 have less freedom in their parametrization, they end up having, somewhat artificially, smaller uncertainties.
The importance of the dijet data is clearly visible when we compare the gluon modifications. Since nCTEQ15 did not include the LHC dijet data and also had more restrictive $Q^2$ cut for DIS data, they have larger uncertainties than what we obtain in EPPS16.
This can also be seen in Figure~\ref{fig:LHC}, where we show a prediction for the dijet forward-to-backward ratio using the nCTEQ15 nuclear PDFs, the uncertainties being significantly larger than those of the CMS data or EPPS16. The DSSZ nPDFs~\cite{deFlorian:2011fp}, which have very little nuclear modifications for gluons, when combined with the CT14 proton PDFs, do not seem to reproduce the dijet data.

\section{Conclusions and outlook}

We have discussed here the advances in the EPPS16 nuclear PDFs, with an emphasis on the impact of the LHC data. While the Z and W data do not yield stringent constraints yet, the included dijet data prove crucial in setting the shape of nuclear gluon modifications. The near future prospects are also good: dijet $R_{\rm pPb}$, sensitive to gluons in a wide $x$ range from $0.8$ down to $10^{-3}$, are being prepared by CMS~\cite{CMS:2016kjd}, while $D^0$ production in LHCb~\cite{Aaij:2017gcy} at forward rapidities can probe gluons at very small $x$, even at $10^{-5}$. In this regard, we are expecting a rapid development of nPDFs.

\vspace{-0.8ex}
\acknowledgments
\vspace{-0.8ex}

We have received funding from the Academy of Finland, Project 297058 of K.J.E.\ and 308301 of H.P.; the European Research Council grant HotLHC ERC-2011-StG-279579; Ministerio de Ciencia e Innovaci\'{o}n of Spain and FEDER, project FPA2014-58293-C2-1-P; and Xunta de Galicia (Conselleria de Educacion) - C.A.S.\ is part of the Strategic Unit AGRUP2015/11. P.P.\ acknowledges the financial support from the Magnus Ehrnrooth Foundation. Part of the computing has been done in T. Lappi's project at CSC, the IT Center for Science in Espoo, Finland.

\end{document}